\documentclass{phbauth}
\usepackage{graphicx}
\usepackage{amssymb}

\begin{document}

\begin{frontmatter}

\title{Perturbation study of the conductance through \\
 a finite Hubbard chain}

\author{Akira Oguri\thanksref{thank1}}

\address{
Department of Material Science, Faculty of Science, Osaka City University, 
Osaka 558-8585, Japan}
\thanks[thank1]{E-mail: oguri@sci.osaka-cu.ac.jp}

%
%

\begin{abstract}
The dc conductance through a Hubbard chain of size $N$ (=1,2,\,3,\,\ldots)
connected to noninteracting leads is studied at $T = 0$ in
an electron-hole symmetric case using a perturbation theory in $U$. 
The result shows a typical even-odd property corresponding to 
a Kondo or Mott-Hubbard physics.
\end{abstract}

\begin{keyword}
quantum transport; electron correlation; Fermi liquid; 
mesoscopic system
\end{keyword}
\end{frontmatter}

Motivated by a current interest in effects of electron 
correlation on the transport through small systems, 
we have examined some theoretical approaches 
\cite{ao6,ao7}.
In this report, using a perturbation approach, 
we study the size ($N$) dependence of the transport 
through a small interacting chain connected to 
semi-infinite leads.

We start with 
the Hamiltonian  $\,\mathcal{H} = \mathcal{H}_0 + \mathcal{H}_I$, 
\begin{eqnarray}
\mathcal{H}_0  &=& 
 - \sum_{\scriptstyle i=-\infty\atop \scriptstyle \sigma}^{+\infty} 
 t_{i}^{\phantom{\dagger}}
  \left (\,  c^{\dagger}_{i+1 \sigma}\, c^{\phantom{\dagger}}_{i \sigma}
     \, + \, c^{\dagger}_{i \sigma}\, c^{\phantom{\dagger}}_{i+1 \sigma}
     \, \right ) 
       \nonumber \\
& &  
 - \mu \sum_{\scriptstyle i=-\infty\atop \scriptstyle \sigma}^{+\infty} 
      n_{i \sigma}^{\phantom{\dagger}}
   +  \sum_{\scriptstyle j=1\atop \scriptstyle \sigma}^{N}  
     \left(\epsilon_0 + {U \over 2} \right)
      n_{j \sigma}^{\phantom{\dagger}}
\;, 
\label{eq:H_0} 
\\
\mathcal{H}_I  &=&  U  \sum_{j=1}^{N} \left[\, 
                 n_{j \uparrow}^{\phantom{\dagger}}
             \,n_{j \downarrow}^{\phantom{\dagger}}
- {1 \over 2}    (   n_{j \uparrow}^{\phantom{\dagger}}
                   + n_{j \downarrow}^{\phantom{\dagger}} \, )
 \right]
\;.
\label{eq:H_I} 
\end{eqnarray}
Here $c^{\dagger}_{i \sigma}$ 
is a creation operator for an electron with spin $\sigma$ at site $i$, 
$n_{i \sigma}^{\phantom{\dagger}}
= c^{\dagger}_{i \sigma}\, c^{\phantom{\dagger}}_{i \sigma}$. 
The hopping matrix element is 
uniform $\,t_i^{\phantom{\dagger}} = t\,$ except 
at the boundaries between the central region and two leads; 
$\,t_0^{\phantom{\dagger}}=v_L^{\phantom{\dagger}}\,$ 
and $\,t_N^{\phantom{\dagger}}=v_R^{\phantom{\dagger}}$.

At $T=0$, 
the dc conductance $g_N^{\phantom{\dagger}}$ can be written 
in term of an inter-boundary element of 
a single-particle Green's function $G_{N 1}(\omega+i0^+)$, 
and is determined by the value at the Fermi level $\omega=0$ as 
$g_N^{\phantom{\dagger}}   = (2 e^2/ h) \, 
 4\, \Gamma_L(0) \Gamma_R(0) \left| G_{N 1}(i0^+) \right|^2$ \cite{ao6}.
Here $\Gamma_{\alpha}(0) =  \pi  D(0)\, v_{\alpha}^2$ with $\alpha=L, R$, 
and $D(0)=\sqrt{4t^2-\mu^2}\,/\,(2 \pi t^2)$. 
%
The assumption made here is the validity of the perturbation theory in $U$.
This seems to be probable for small $N$.
In that case,  
the self-energy due to $\mathcal{H}_I$ has a property 
$\mbox{Im}\,\Sigma_{jj'} (i0^+)=0$ at $T=0$  \cite{LangerAmbegaokar},
and $g_N^{\phantom{\dagger}}$ can be obtained through 
a scattering problem of free quasi-particles \cite{ao7}.

In this report, we consider an electron-hole symmetric case 
taking the parameters to be $\mu=0$ and $\epsilon_0+U/2=0$.
If the system has an additional inversion symmetry 
$v_L^{\phantom{\dagger}} = v_R^{\phantom{\dagger}}$, 
it can be shown that a perfect transmission 
occurs $\,g_{N}^{\phantom{\dagger}} \equiv 2e^2/h$ 
for odd $N$ ($=2M+1$) independent of the values of $\,U$ and $M$ \cite{ao7}. 
This is caused by the Kondo resonance 
appearing at the Fermi level for odd $N$.

\begin{figure}[t]
\begin{flushleft}\leavevmode
\vspace{-0.45cm}
\includegraphics[width=1.25\linewidth]{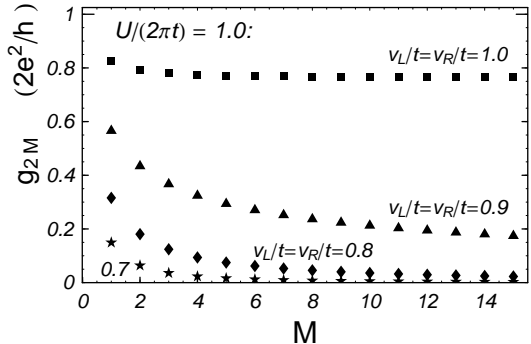}
\vspace{-8.6cm}
\caption{\ 
 $g_{N}^{\protect\phantom{\dagger}}$ vs $\,M\,$  for even $N$ ($=2M$): 
$\,U/(2 \pi t) = 1.0$, and 
$v_L^{\protect\phantom{\dagger}}/t=v_R^{\protect\phantom{\dagger}}/t
=\,0.7(\bigstar) 
,\,0.8(\blacklozenge) 
,\,0.9(\blacktriangle), \,1.0(\blacksquare)$. 
}
\label{fig:g_N_symm}
\end{flushleft}
\end{figure}

\begin{figure}[t]
\begin{center}\leavevmode
\vspace{-0.40cm}
\includegraphics[width=1.2\linewidth]{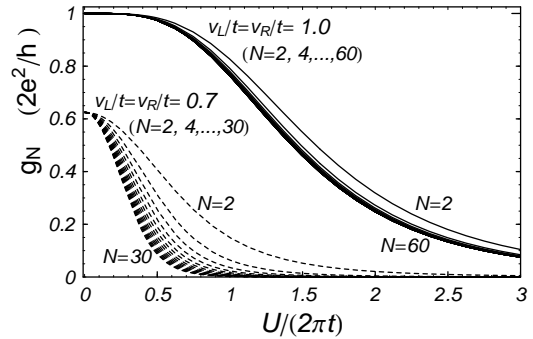}
\vspace{-8.05cm}
\caption{\ $g_N^{\protect \phantom{\dagger}}$ vs $\,U$ for
 even $N$ ($=2,\,4,\,\ldots$):
$\,v_L^{\protect\phantom{\dagger}}/t=v_R^{\protect\phantom{\dagger}}/t =$
 $0.7$ (dashed lines),   $1.0$ (solid lines). 
15+30 lines are plotted. 
}
\label{fig:g_N_U}
\end{center}
\end{figure}

On the other hand, for even $N$, 
we evaluate the self-energy $\Sigma_{jj'}(i0^+)$ 
within the second order in $U$ at $T=0$, 
and then obtain $G_{N 1}(i0^+)$
solving the Dyson equation in the real space \cite{ao7}.
In Fig.\ \ref{fig:g_N_symm},    
$\,g_N^{\phantom{\dagger}}$ for even $N$ ($=2M$)
is plotted vs $M$ for several values of 
$v_L^{\phantom{\dagger}}$ ($= v_R^{\phantom{\dagger}}$),  
where $U/(2\pi t) =1.0$. 
The dc conductance decreases with increasing the size $2M$. 
This behavior can be regarded as a tendency toward a Mott-Hubbard insulator, 
and it is pronounced for larger $U$.
In Fig.\ \ref{fig:g_N_U}, 
$\,g_N^{\phantom{\dagger}}$ for even $N$ ($=2,\,4,\,\ldots$) is 
plotted vs $U$. 
The value of $\,g_{2M}^{\phantom{\dagger}}$  decreases with increasing $U$. 
When $v_L^{\phantom{\dagger}}$ (or $v_R^{\phantom{\dagger}}$) is 
smaller than $t$, 
the reduction of $g_{2M}^{\phantom{\dagger}}$ is proportional to $\,U^2$ 
at $\,U/(2\pi t) \ll 1$. 
As it is seen in the plots  
for $v_L^{\phantom{\dagger}} = v_R^{\phantom{\dagger}} = 0.7\,t$ 
(dashed lines), the peak structure 
in the $U$ dependence becomes sharp for large $M$, 
and in the limit $M \to \infty$ the peak seems to vanish 
leaving the value at a singular point $U=0$ unchanged.
In contrast, 
in the case of $v_L^{\phantom{\dagger}} 
= v_R^{\phantom{\dagger}} = t$ (solid lines), 
the reduction of $\,g_{2M}^{\phantom{\dagger}}$  is 
proportional to $\,U^4$ at $\,U/(2\pi t) \ll 1$,  
and $g_{2M}^{\phantom{\dagger}}$ seems to be finite 
in the limit of large $M$. 
However, in order to verify this behavior for large $M$,
the contributions of the higher-order terms should be examined 
because the unperturbed Hamiltonian $\mathcal{H}_0$ 
has a translational invariance accidentally in this case.

Qualitatively, the even-odd property seems to be understood from 
that in the unperturbed system, 
especially from the level structure of the isolated chain.
For odd $N$, there is a semi-occupied 
one-particle state at the Fermi level $\omega=0$, 
and thus a doublet ground state is realized.
When the leads are connected, 
the ground state is replaced by a Kondo singlet state 
and contributes to the tunneling. 
On the other hand, for even $N$, the Fermi level lies between 
the highest occupied level and the lowest unoccupied level, 
and thus a finite energy corresponding to the gap is necessary 
to excite the electrons.
Although the levels are broadened by the coupling with the leads,
the even-odd property is determined by  
whether there exists a zero-energy excitation or not. 
At finite temperatures, a characteristic energy scale of 
the Kondo or Mott-Hubbard physics will play an important role.

We expect that the model and approach used here 
can be applied to a series of quantum dots 
or a quantum wire of nanometer scale.

We thank H.\ Ishii for valuable discussions.

\end{document}